\pdfoutput=1
\documentclass[aps,prd,reprint,superscriptaddress,longbibliography,nofootinbib]{revtex4-1}

\usepackage{graphicx}
\usepackage{amsmath}
\usepackage{amssymb}
\usepackage{amsfonts}
\usepackage{dcolumn}
\usepackage{dsfont}
\usepackage{latexsym}
\usepackage{rotating}
\usepackage{color}
\usepackage{latexsym}
\usepackage{bbm}
\usepackage{subfigure}
\usepackage{float}
\usepackage{epsfig}
\usepackage{psfrag}
\usepackage{natbib, hyperref}
\usepackage{bm}
\usepackage{amsthm}
\usepackage{eucal}
\usepackage{mathrsfs}
\usepackage{url}
\usepackage{braket}
\usepackage{ulem}

\usepackage{calligra}
\usepackage[T1]{fontenc}
\usepackage[utf8]{inputenc}
\usepackage{newunicodechar}
\usepackage{marvosym}
\usepackage[absolute]{textpos}
\usepackage[cal=cm]{mathalfa}

\usepackage{color} 

\usepackage{hyperref}
\hypersetup{
colorlinks=true,final=true,
        linkcolor=blue,
        citecolor=blue,
        filecolor=blue,
        urlcolor=blue,
}
\usepackage{amssymb,amsmath,amsthm,enumitem}

\begin{document}
\setlength{\abovedisplayskip}{4pt}
\setlength{\belowdisplayskip}{4pt}

\title{Magnetic Switching in Weyl Semimetal-Superconductor Heterostructures}
\thanks{Copyright notice: This manuscript has been authored by UTBattelle, LLC under Contract No. DE-AC05-00OR22725 with the U.S. Department of Energy. The United States Government retains and the publisher, by accepting the article for publication, acknowledges that the United States Government retains a non-exclusive, paid-up, irrevocable, world-wide license to publish or reproduce the published form of this manuscript, or allow others to do so, for United States Government purposes. The Department of Energy will provide public access to these results of federally sponsored research in accordance with the DOE Public Access Plan (http://energy.gov/downloads/doepublic-access-plan).}
\author{Narayan Mohanta}
\email{E-mail: mohantan@ornl.gov}
\affiliation{Material Science and Technology Division, Oak Ridge National Laboratory, Oak Ridge, TN 37831, USA}
\author{A. Taraphder}
\affiliation{Department of Physics and Center for Theoretical Studies, Indian Institute of Technology Kharagpur, W.B. 721302, India}
\author{Elbio Dagotto}
\affiliation{Material Science and Technology Division, Oak Ridge National Laboratory, Oak Ridge, TN 37831, USA}
\affiliation{Department of Physics and Astronomy, The University of Tennessee, Knoxville, TN 37996, USA}
\author{Satoshi Okamoto}
\affiliation{Material Science and Technology Division, Oak Ridge National Laboratory, Oak Ridge, TN 37831, USA}

\begin{abstract}
We present a new switching mechanism that utilizes the proximity coupling between the surface spin texture of a Weyl semimetal and a superconductor, in a Weyl semimetal/superconductor/Weyl semimetal trilayer heterostructure. We show that 
the superconductivity in the middle layer can be fully suppressed by the surface spin texture of the Weyl semimetals in the presence of an external magnetic field, but it can be recovered again by only changing the field direction. The restoration of the middle-layer superconductivity indicates a sharp transition to a low-resistance state. This sharp switching effect, realizable using a Weyl semimetal because of its strong spin-momentum locking and surface spin polarization, is a promising avenue for novel superconducting spin-valve applications.
\end{abstract}

\maketitle

\section{Introduction}
\vspace{-4mm}
A three-dimensional Weyl semimetal (WSM) is strikingly different from other classes of materials with non-trivial band topology because of the presence of surface Fermi arcs~\cite{Muff_PRL2015,Sun_PRB2015} which prompted the observation of a plethora of intriguing properties, such as quantum oscillations in magnetoresistance~\cite{Potter_NComm2014,Moll_Nature2016,Moll_NComm2016} and chiral magnetic effect~\cite{Nielsen_PLB1983,Son_PRB2013,Kim_PRL2013,Burkov_PRL2014,Li_NComm2015,Huang_PRX2015,Li_NPhys2016,Li_PRB2016,Wang_PRB2016,Klotz_PRB2016,Baireuther_NJP2016,Taguchi_PRB2016,Lin_PRB2018}. The Fermi arcs exhibit a spin texture with a strong spin-momentum locking which leads to a spin polarization (up to 80$\%$ for TaAs) on the surface of the WSMs~\cite{Li_NatCommu2018,Xu_PRL2016,Feng_PRB2016}. By introducing a superconducting gap in the WSM, the superconductor (SC) inherits the non-trivial topology of the electronic structure of the WSM, giving rise to unconventional properties such as finite-momentum pairing and Majorana zero modes~\cite{PhysRevB.86.214514,PhysRevB.92.035153,PhysRevB.90.045130,PhysRevB.86.054504,PhysRevB.93.201105,Baireuther_NJP2017,Qi2016,PhysRevB.94.224512,10.1063.1.4947433,WANG2017425,Guguchia2017,Bachmanne1602983,Aggarwal_NComm2017}. At a WSM/SC interface, superconductivity is proposed to be induced by the proximity effect inside the WSM near the interface~\cite{PhysRevB.90.195430}. Despite a few studies on the proximity effect of the superconductivity inside the WSM in a WSM/SC interface, the {\it inverse} proximity effect of the WSM surface magnetization on the superconductivity remains unexplored.
\begin{figure}[t]
\begin{center}
\epsfig{file=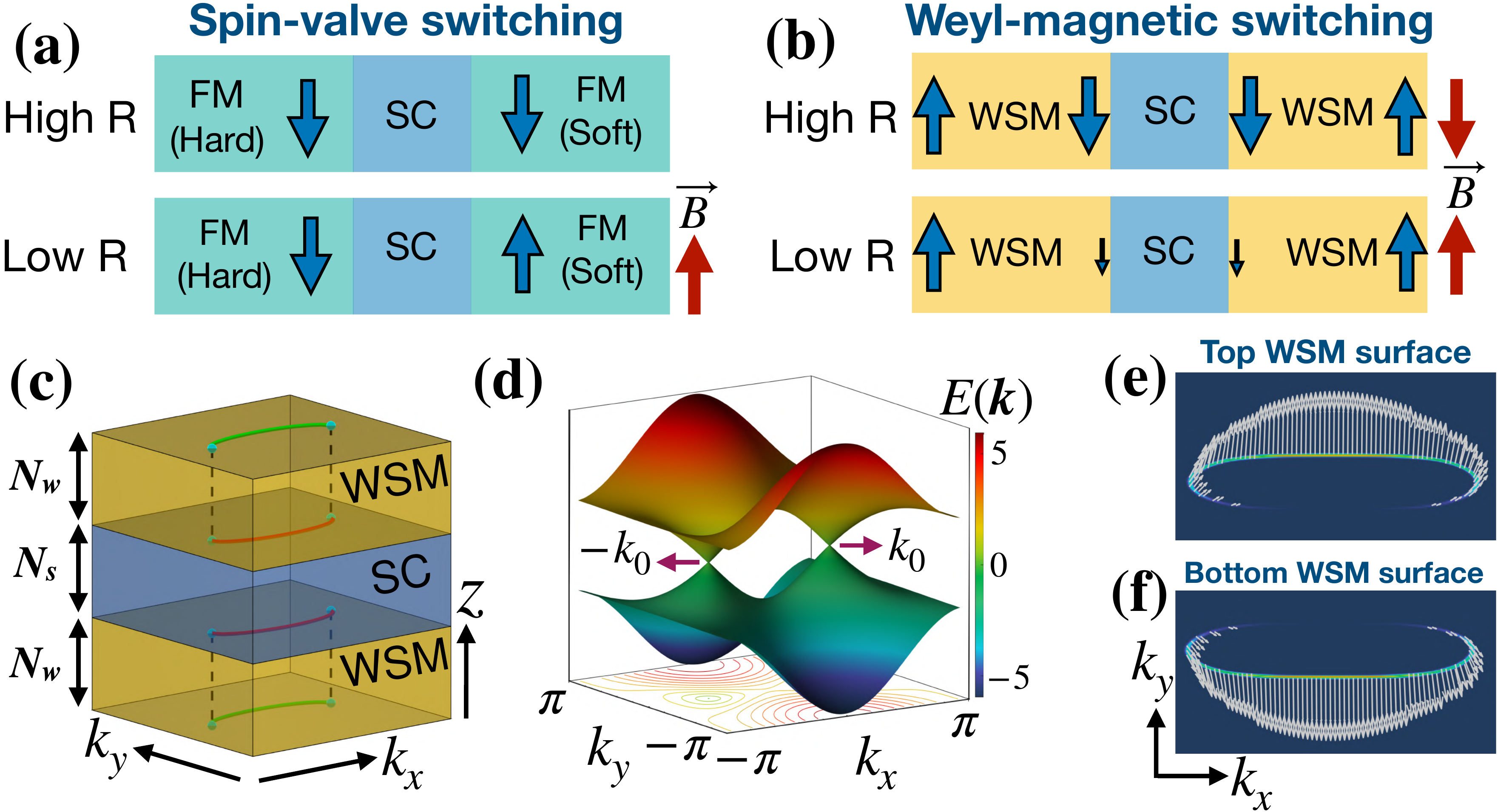,trim=0.0in 0.0in 0.0in 0.0in,clip=false, width=84mm}
\caption{(a) Standard spin-valve switching mechanism involving a FM/SC/FM trilayer. The blue arrows denote the magnetization in the FMs. (b) Proposed switching mechanism using a WSM/SC/WSM trilayer. The blue arrows denote the net surface magnetization. (c) Schematic of the considered trilayer heterostructure. The red and the green lines at the surfaces of the WSMs denote the conjugate Fermi arcs. (d) Bulk bands of a WSM with broken time-reversal symmetry, showing the pair of the Weyl nodes at $(\pm k_0,\!0,\!0)$. $E(\mathbf{k})$  are the eigenvalues of Hamiltonian~(\ref{H_wsm}) at $k_z$=$0$ (in units of $t$, a parameter mentioned in the text). (e)-(f) The computed Fermi arcs (the momentum-resolved density of states at energy $E(\mathbf{k})$=$-0.3t$) and spin textures (white arrows) on the opposite surfaces of an isolated WSM slab of thickness $N_w$=$15$. }
\label{fig1}
\vspace{-6mm}
\end{center}
\end{figure}

In this work, we demonstrate a new switching effect induced by the interplay between superconductivity and magnetization at a WSM/SC interface, in the presence of an external magnetic field. We consider a trilayer heterostructure, with an $s$-wave SC sandwiched between two WSM slabs in such a way that the top and the bottom WSMs have opposite alignments of the Fermi arcs. In this geometry, the WSM surfaces, on both sides of the SC, have a net spin polarization {\it in the same direction}. The superconductivity in the middle SC layer is suppressed completely by the joint pair-breaking effect of the WSM surface magnetization and the magnetic field. However, by rotating the field, the net magnetization of the WSM surfaces drops to a smaller value and the fully-suppressed superconductivity restores again, indicating a sharp transition to the zero-resistance state in the middle SC layer. Such a magnetic switching has the potential to advance the spin-valve applications that at present employ 
trilayer heterostructures of ferromagnets (FMs) and SCs~\cite{Speriosu_PRB1993,Moodera_PRL1998,Buzdin_EPL1999,Tagirov_PRL1999,Takahashi_PRL1999,Baladi_PRB2001,Gu_PRL2002,Potenza_PRL2005,Pena_PRL2005,Rusanov_PRB2006,Yanson_NLett2007,Zhu_PRL2009,PhysRevLett.105.256804,Simoni_NLett2018,Halterman_PRL2013,Majidi_PRB2014,Alidoust_PRB2018}.

In the state-of-the-art spin-valve switching devices, involving a FM(hard)/SC/FM(soft) trilayer, as shown in  Fig.~\ref{fig1}(a), the resistance  $R_p$, in the parallel polarization of the top and bottom FMs near the superconducting transition temperature $T_c$, is larger than the resistance $R_{ap}$ in the anti-parallel case. The magnetic field flips the polarization in the soft FM and produces a transition to a low-resistance state. In our proposed WSM-based geometry, shown in Figs.~\ref{fig1}(b) and (c), the net magnetization of the WSM surfaces on both sides of the SC drops to a smaller value when the magnetic field is applied opposite to the spin polarization direction. The field rotation drives the middle SC to the zero-resistance state, implying a low resistance in the WSM/SC/WSM trilayer.

\section{Model and method}
\vspace{-4mm}
We consider a type-I WSM with broken time-reversal symmetry and two Weyl points, described by the following Hamiltonian~\cite{PhysRevB.95.075133}
\begin{align}
\mathcal{H}_{w}\!=\!&\sum_{\mathbf{k}}\!-\Big[ \{ m(2\!-\!\cos{k_y}\!-\!\cos{k_z}) \nonumber 
\!+\!2t_x(\cos{k_x}\! -\!\cos{k_0}) \} \sigma_{x} \nonumber \\
&-(2t \sin{k_y}) \sigma_{y}\!-\!(2t \sin{k_z}) \sigma_{z}\!-\!\mu_w \mathcal{I} \Big],
\label{H_wsm}
\end{align}
written in the basis $\big( c_{\mathbf{k},\uparrow}, c_{\mathbf{k},\downarrow}\big)^{T}$, $m$, $t_x$, and $t$ describe the bulk band dispersion, $\mu_w$ is the chemical potential, $\boldsymbol{\sigma}$$\equiv$$( \sigma_x\!,\sigma_y,\!\sigma_z )$ are the Pauli matrices acting on the pseudo spins, $\mathcal{I}$ is the $2\times2$ identity matrix, and $\mathbf{k}$$\equiv$$( k_x,\! k_y,\! k_z)$ denotes the momentum inside the WSM. The Weyl nodes are located at $(\pm k_0,\!0,\!0)$ as depicted in Fig.~\ref{fig1}(d). 
In Fig.~\ref{fig1}(e) and (f), we show the Fermi arcs and the spin texture on the opposite surfaces of an isolated WSM slab~\cite{SM}. The parameters $t$=$1$,~$m$=$2t$,~$k_0$=$\pi/2$,~$t_{x}$=$t$, used in Ref.~\cite{PhysRevB.95.075133}, are kept fixed, with no qualitative change in our conclusions for other choices.

We consider BCS-type electron pairing in the middle SC slab, expressed by the following Hamiltonian~\cite{SM} 
\begin{align}
\mathcal{H}_{s}\!=\!&- \sum_{\mathbf{k},\sigma} \Big[ 2t_s(\cos{k_x}\!+\!\cos{k_y}\!+\!\cos{k_z})\!+\!\mu_s\Big] d_{\mathbf{k},\sigma}^{\dagger} d_{\mathbf{k},\sigma} \nonumber \\
&+\sum_{\mathbf{k}} (U_0\Delta_s d_{\mathbf{k},\uparrow}^{\dagger}d_{\mathbf{-k},\downarrow}^{\dagger}+H.c.) + N U_0 \Delta_{s}^{2},
\end{align}
where $t_s$ is the hopping amplitude, $\mu_s$ the chemical potential, $\Delta_{s}$=$\langle d_{\mathbf{k},\uparrow} d_{\mathbf{-k},\downarrow} \rangle$ the $s$-wave pairing amplitude in the SC, $U_0$ the pairwise attractive interaction strength, and $N$ is the total number of momenta in the Brillouin zone.  We set $t_s$=$1.5t$ and $U_0$=$-t_s$ everywhere, without losing generality.
\begin{figure}[t]
\begin{center}
\epsfig{file=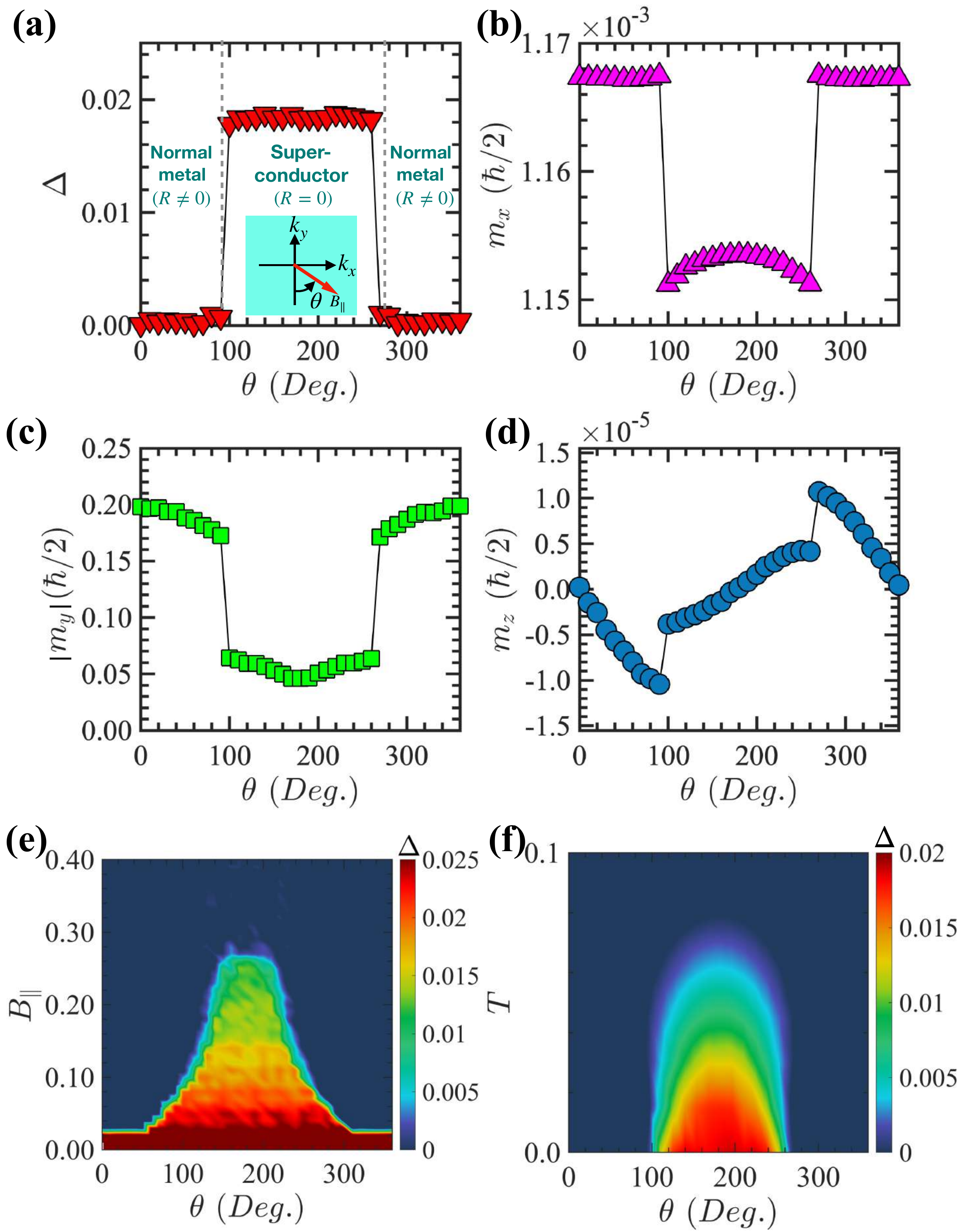,trim=0.0in 0.0in 0.0in 0.0in,clip=false, width=84mm}
\caption{(a) Superconducting pairing amplitude $\Delta$, averaged over all layers of the WSM/SC/WSM heterostructure, plotted as a function of the magnetic-field angle $\theta$ which is measured with respect to the $-k_y$ direction, as shown in the inset. (b)-(d) $\theta$-variation of the magnetization components $m_x$, $m_y$, $m_z$ at the bottom surface of the top WSM slab. The magnetic field and the temperature in plots (a)-(d) are $B_{\parallel}$=$0.1t$ and $T$=$0$, respectively. (e), (f) Variation of $\Delta$ (number in the colorbar) with the angle $\theta$ and the magnetic-field amplitude $B_{\parallel}$ (at $T$=$0$) in (e), and temperature $T$ (at $B_{\parallel}$=$0.1t$) in (f). Parameters used are $N_w$=$15$, $N_s$=$15$,~$t_{tun}$=$t$,~$\mu_{w}$=$-0.3t$ and $\mu_{s}$=$0.1t$.}
\label{theta_var}
\vspace{-6mm}
\end{center}
\end{figure}

The Hamiltonian for the heterostructure is constructed by transforming the Hamiltonians $\mathcal{H}_{w}$ and $\mathcal{H}_{s}$ into the slab geometry~\cite{SM,Mohanta_SciRep2017}, with $N_s$ layers in the SC slab and $N_w$ layers in both (top and bottom) WSM slabs. Open boundary conditions are imposed only along the stacking direction (the $z$ axis), so that $k_x$ and $k_y$ are good quantum numbers for the description in each layer.

The coupling between the WSM and the SC slabs at the WSM/SC interfaces is described by the following tunneling Hamiltonian
\begin{align}
\mathcal{H}_{tun}\!=\!- \sum_{\mathbf{k^{\parallel}},l_z,\sigma} (t_{tun} c_{\mathbf{k^{\parallel}},\sigma,l_z}^{\dagger} d_{\mathbf{k^{\parallel}},\sigma,l_z+1}+H.c.),
\end{align}
where $l_z$ is the vertical layer index at the interface, $t_{tun}$ is the tunneling amplitude between the WSMs and the SC and $\mathbf{k^{\parallel}}$$\equiv$$(k_x,\! k_y)$ is the momentum in the $k_x$-$k_y$ plane. 

The external magnetic field, applied uniformly in all the layers of the heterostructure, is included as a Zeeman exchange coupling to the pseudo spins
\begin{align}
\mathcal{H}_z\!=\!\sum_{\mathbf{k^{\parallel}},l_z,\sigma,\sigma^{\prime}} (\mathbf{B} \cdot \boldsymbol{\sigma}_{\sigma \sigma^{\prime}}) c_{\mathbf{k^{\parallel}},\sigma,l_z}^{\dagger} c_{\mathbf{k^{\parallel}},\sigma^{\prime},l_z}, 
\end{align}
where  $\mathbf{B}$=$(B_{\parallel}\cos{\theta'},B_{\parallel}\sin{\theta'},0)$, $B_{\parallel}$ is the strength of the magnetic field, $\theta'$=$3\pi/2+\theta$ and $\theta$ is the field angle as shown in Fig.~\ref{theta_var}(a). Henceforth, all energies are expressed in units of the parameter $t$ defined in Eq.~(\ref{H_wsm}).\\
\indent The pairing amplitude in each layer $l_z$ inside the SC slab is obtained self-consistently using $\Delta_{s,l_z}$=$\langle d_{\mathbf{k^{\parallel}},l_z,\uparrow} d_{\mathbf{-k^{\parallel}},l_z,\downarrow} \rangle$~\cite{SM}. The  proximity-induced $s$-wave pairing amplitude inside the WSM slab is calculated via $\Delta_{w,l_z}$=$\langle c_{\mathbf{k^{\parallel}},l_z,\uparrow} c_{\mathbf{-k^{\parallel}},l_z,\downarrow} \rangle$. Although a small finite-momentum pairing and a triplet pairing could also be induced in the WSM, we focus only on singlet pairing at zero center-of-mass momentum which is the dominant pairing channel and relevant to the present study. The average pairing amplitude for the heterostructure is then obtained via $\Delta$=$(1/(N_s+2N_w))\sum_{l_z}(\Delta_{s,l_z}+\Delta_{w,l_z})$. The spin texture at a layer $l_z$ ($m_{\mathbf{k}^{\parallel},x}$, $m_{\mathbf{k}^{\parallel},y}$, $m_{\mathbf{k}^{\parallel},z}$) is obtained by computing the spin-expectation values at each momentum $\mathbf{k}^{\parallel}$ and the average magnetization components ($m_{x}$, $m_{y}$, $m_{z}$) are obtained by taking average in the two-dimensional Brillouin zone~\cite{SM}.

\section{Origin of the magnetic switching}
\vspace{-4mm}
The main concept introduced here, the magnetic switching at the WSM/SC interfaces, involves the strong interplay between superconductivity and the WSM surface magnetization; the latter competes against the electron pairing near the interfaces. The magnetic field $B_{\parallel}$=$0.1t$ (<$B_{\parallel}^{c}$$\approx$$0.3t$, the critical field for the SC), applied along $\theta$=$0^{\circ}$, cooperates with the WSM surface magnetization to {\it completely} suppress superconductivity. Remarkably, by changing the field direction, the average pairing amplitude $\Delta$ re-emerges and jumps from nearly zero to a finite value at $\theta$$\approx$$100^{\circ}$. The order parameter $\Delta$ remains almost constant within a range $100^{\circ}$$\lesssim$$\theta$$\lesssim$$260^{\circ}$, as shown in Fig.~\ref{theta_var}(a). This sudden restoration of the superconductivity, only by changing the field angle, implies a switching from a finite-resistance to a low-resistance state. {\it This magnetic field-driven `switching' is the main finding of this paper.} 

To understand the origin of this switching effect, we explore the $\theta$-dependence of the magnetization components. Interestingly, we find similar discontinuous jumps in the magnetization components at the WSM terminating layers, adjacent to the SC slab, as shown in Figs.~\ref{theta_var}(b)-(d). The in-plane magnetization components $m_x$ and $m_y$ drop to relatively smaller values within the same range $100^{\circ}$$\lesssim$$\theta$$\lesssim$$260^{\circ}$. The out-of-plane component $m_z$ also undergoes sharp transitions at the critical field angles $\theta_{c1}$$\approx$$100^{\circ}$ and $\theta_{c2}$$\approx$$260^{\circ}$, as shown in Fig.~\ref{theta_var}(d). This finding underlines the strong interplay between the magnetism in the WSM surface and the superconductivity near the WSM/SC interfaces. The $\theta$-variation of $\Delta$ with the field amplitude $B_{\parallel}$, in Fig.~\ref{theta_var}(e), suggests that there is a lower and an upper critical field and 
the switching occurs in between. The lower critical field is the required field to fully suppress the superconductivity at certain ranges of $\theta$ values, while the upper critical field is the largest field above which the reappearance of superconductivity is unanticipated. Temperature variation of $\Delta$, in Fig.~\ref{theta_var}(f), suggests that the switching effect appears predominantly at low temperatures, below the superconducting $T_c$ of the middle SC. 
\begin{figure}[t]
\begin{center}
\epsfig{file=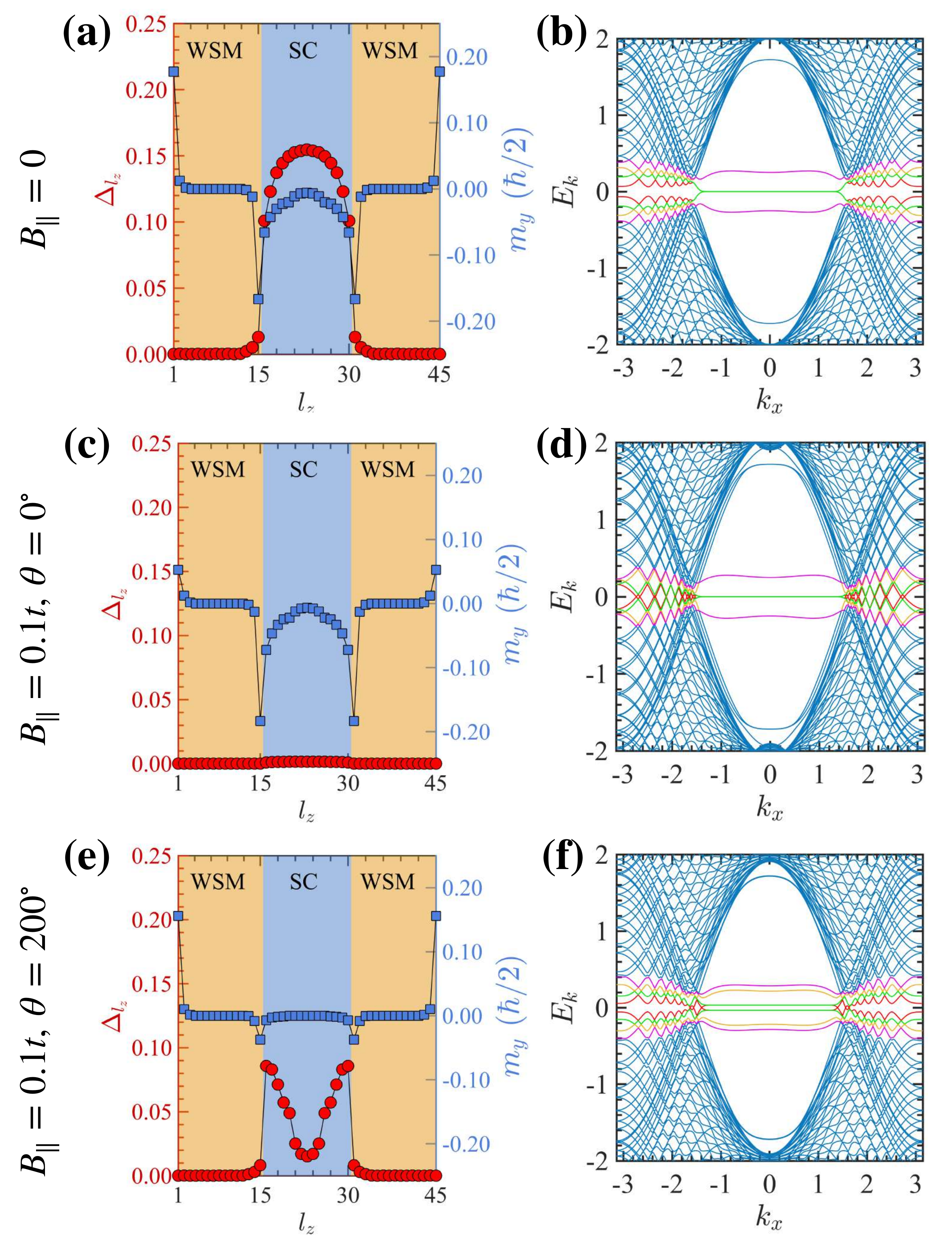,trim=0.0in 0.05in 0.0in 0.0in,clip=true, width=84mm}
\caption{(a),(c),(e) Variation of the superconducting pairing amplitude $\Delta_{l_z}$ (left vertical axis) and  $y$ component of magnetization $m_y$ (right vertical axis) with the layer index $l_z$ at (a) magnetic field $B_{\parallel}$=$0$, (c) $B_{\parallel}$=$0.1t$, $\theta$=$0^{\circ}$, and (e) $B_{\parallel}$=$0.1t$, $\theta$=$200^{\circ}$. (b), (d), and (f) show the corresponding energy spectrum of the WSM/SC/WSM heterostructure at $k_y$=$0$. The energy levels in red, green, yellow and magenta originate primarily from the surfaces of the WSMs. Parameters are the same as in Fig.~\ref{theta_var}.}
\label{bands}
\vspace{-6mm}
\end{center}
\end{figure}

To explore the interplay between the WSM surface magnetization and the superconductivity, we plot the layer-resolved pairing amplitude $\Delta_{l_z}$(=$\Delta_{s,l_z}~{\rm{or}}~\Delta_{w,l_z}$), and the dominant magnetization component $m_y$, in Figs.~\ref{bands}(a), (c) and (e), at different field scenarios and the corresponding spectrum in Figs.~\ref{bands}(b), (d) and (f). At $B_{\parallel}$=$0$, a small pairing amplitude is induced inside the WSMs, by proximity effect, but this induced superconductivity is limited to the near 
vicinity of the interfaces~\cite{PhysRevB.90.195430}. Conversely, the pairing amplitude inside the SC is also suppressed near the interfaces since the WSM surface magnetism penetrates inside the SC and affects the superconductivity. At a finite magnetic field $B_{\parallel}$=$0.1t$, applied along $\theta$=$0^{\circ}$, the superconductivity is completely suppressed, as shown in Fig.~\ref{bands}(c), by the combined pair-breaking effect of the magnetic field and the WSM surface magnetization. However, when the field is applied along $\theta$=$200^{\circ}$, i.e. opposite to the WSM surface magnetization, the net magnetization of the WSM surface drops drastically which enables the reappearance of the superconductivity inside the SC, as depicted in Fig.~\ref{bands}(e). Similar reappearance of the superconductivity takes place within the angle range $100^{\circ}$$\lesssim$$\theta$$\lesssim$$260^{\circ}$. The profile of $\Delta_{l_z}$ remains nearly unchanged within this range of $\theta$~\cite{SM}. The sharp transitions, at the critical angles, is possible in a WSM because of its strong spin-momentum locking which makes the spin texture of the Fermi arcs quite robust against the external magnetic field, applied at an angle outside this angle range. Within this range, the magnetic field overcomes the spin-momentum locking and significantly reduces the in-plane magnetization components~\cite{SM}.
\begin{figure}[t]
\begin{center}
\epsfig{file=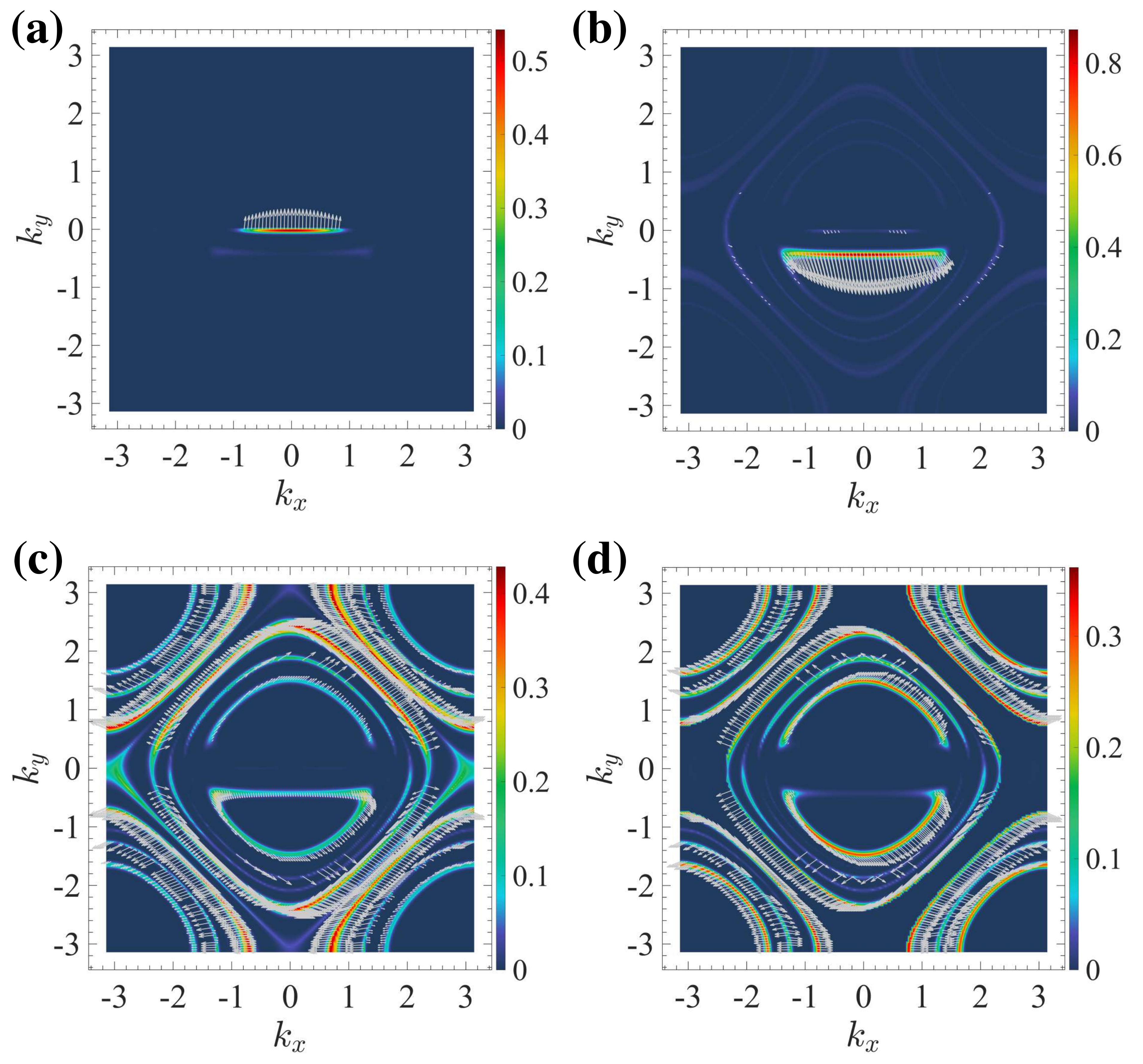,trim=0.0in 0.15in 0.0in 0.0in,clip=true, width=86mm}
\caption{Momentum-resolved density of states $A(\mathbf{k^{\parallel}}\!,\!E)$ (colorbar) and spin texture (arrows show the in-plane spin-expectation values $m_{\mathbf{k}^{\parallel},x}$ and $m_{\mathbf{k}^{\parallel},y}$) at energy $E$=$-0.3t$ at zero magnetic field and different layers of the WSM/SC/WSM heterostructure. (a) $l_z$=$1$ (top layer of the top WSM), (b) $l_z$=$15$ (bottom layer of the top WSM), (c) $l_z$=$16$ (top layer of the SC), (d) $l_z$=$18$ (inside the SC but close to the top WSM/SC interface). Parameters are the same as in Fig.~\ref{theta_var}.}
\label{texture}
\vspace{-6mm}
\end{center}
\end{figure}
\begin{figure*}[t]
\begin{center}
\epsfig{file=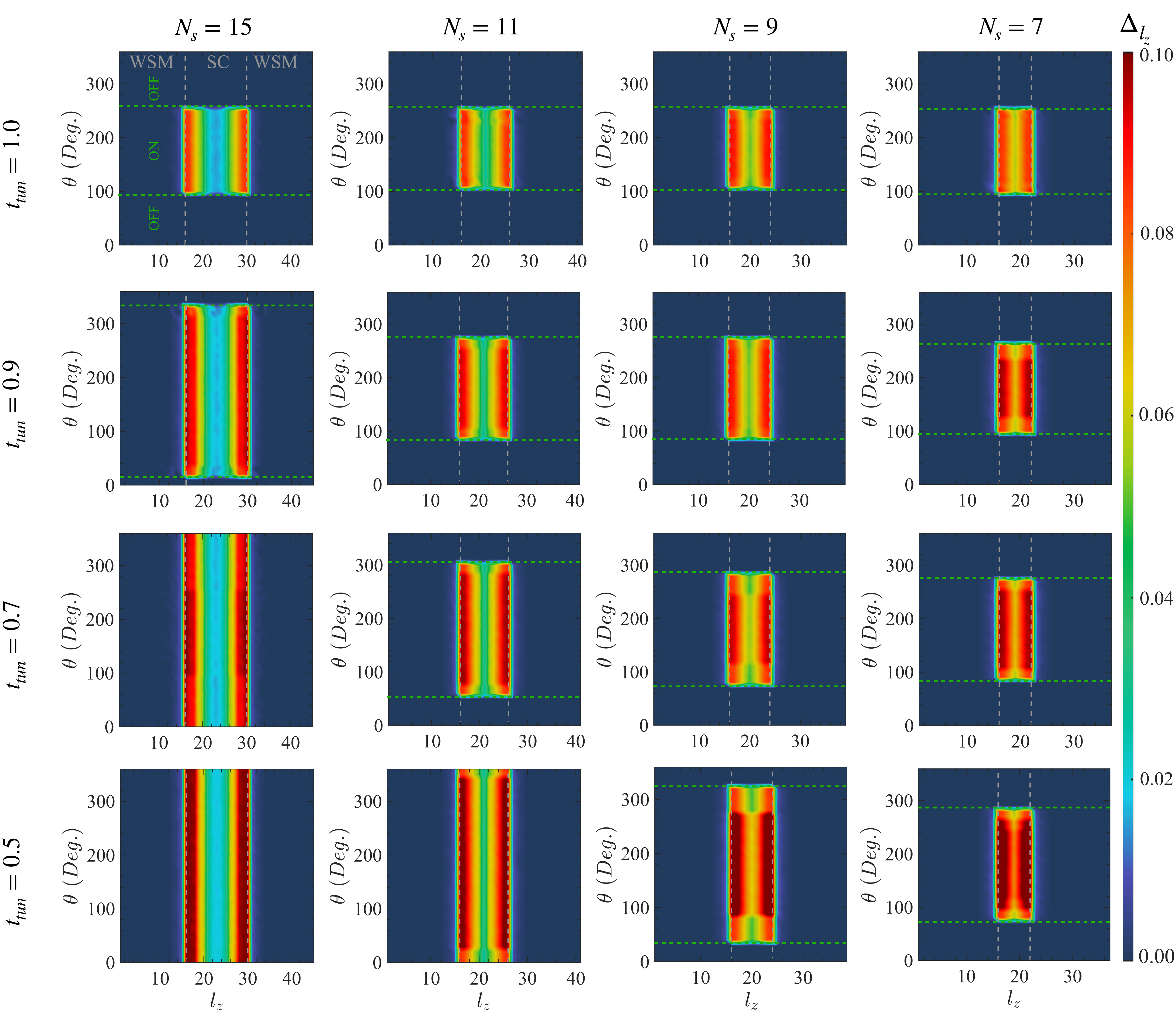,trim=0.0in 0.15in 0.0in 0.0in,clip=false, width=160mm}
\caption{Profiles of the layer-resolved pairing amplitude $\Delta_{l_z}$ (in the $l_z$ - $\theta$ plane) at different hopping energies $t_{tun}$ and the SC layer thicknesses $N_s$. The applied magnetic field is $B$=$0.1t$ and other parameters are the same as in Fig.~\ref{theta_var}. The ON (OFF) region in the top-left panel shows the magnetic-field angle range within which the middle SC region is superconducting (non-superconducting), implying a low-resistance state in the trilayer heterostructure.}
\label{fig5}
\vspace{-6mm}
\end{center}
\end{figure*}

The energy spectrum of the heterostructure at $B_{\parallel}$=$0$, in Fig.~\ref{bands}(b), shows that two pairs of energy levels (green and red) exhibit flat zero-energy states within the Fermi-arc region ($|k_x|$$\lesssim$$k_0$) while two other pairs (magenta and yellow) are gapped out at finite energies. These low-energy bands originate from the hybridization between the surface states in the WSMs and the SC slabs. The spectrum changes when a magnetic field $B_{\parallel}$=$0.1t$ is applied along $\theta$=$0^{\circ}$, as shown in Fig.~\ref{bands}(d). Particularly, outside the flat-band region ($|k_x|$$>$$k_0$), the red and the green bands do not have an energy gap at $\theta$=$0^{\circ}$ (Fig.~\ref{bands}(d)) and a finite energy gap at $\theta$=$200^{\circ}$ (Fig.~\ref{bands}(f)). These two angles belong to the two different states \textit{viz.} the suppressed and recovered superconducting states, shown before in Fig.~\ref{theta_var}(a). Evidently, the electronic spectra also reveals a signature of these different states.

To gain insight on the critical field angles $\theta_{c1}$ and $\theta_{c2}$, we calculated the spin texture and the momentum-resolved density of states $A(\mathbf{k^{\parallel}}\!,\!E)$=$\sum_{\sigma,\lambda}|c_{{\mathbf{k}}^{\parallel},\sigma,l_z}|^2\delta(E$-$E_{\mathbf{k^{\parallel}} \lambda})$+$|c_{-{\mathbf{k}}^{\parallel},\sigma,l_z}^{\dagger}|^2\delta(E$+$E_{-\mathbf{k^{\parallel}} \lambda})$  at different layers of the heterostructure ($c_{{\mathbf{k}}^{\parallel},\sigma,l_z}$ is replaced by $d_{{\mathbf{k}}^{\parallel},\sigma,l_z}$ for the SC layers). In Fig.~\ref{texture}, we show the spin texture  at four different layers at $B_{\parallel}$=$0$. The top surface of the top WSM ($l_z$=1) reveals a Fermi arc and spin texture [Fig.~\ref{texture}(a)]  at energy $E$=$-0.3t$(=$\mu_w$), similar to the case of an isolated WSM slab, shown earlier in Fig.~\ref{fig1}(e). On the other hand, Fig.~\ref{texture}(d), shows the constant-energy contours and spin texture at a SC layer ($l_z$=18), slightly below the top WSM/SC interface, at energy $E$=$-0.3t$ ($<\mu_s$) showing the electron-like and hole-like contours. Notably, two electron-like contours around $\mathbf{k}^{\parallel}$=$\mathbf{0}$ are disconnected along the $k_y$=$0$ direction, influenced by the Fermi arc in the neighboring WSM layer ($l_z$=15). The proximity effect of the Fermi arc is prominent at the top SC layer ($l_z$=16) [in Fig.~\ref{texture}(c)] which also develops a Fermi arc in the central disjoint electron-like contour with its spin texture originating from the adjacent WSM surface [in Fig.~\ref{texture}(b)]. The spin texture at the top WSM/SC interface (both $l_z$=15 and $l_z$=16) results in a net spin polarization, as shown in Fig.~\ref{texture}(b) and (c). This spin polarization suppresses the superconductivity near the WSM/SC interfaces and drives the `switching' transitions.

\section{Discussion}
\vspace{-4mm}
Here, we discuss briefly the robustness of the proposed switching effect against different parameter choices in our model, the most relevant one being the hopping energy $t_{tun}$ between the WSM and the SC. The results presented until now were obtained with  $t_{tun}$=$t$ at SC-layer thickness $N_s$=$15$. In Fig.~\ref{fig5}, we show the layer-resolved pairing amplitude $\Delta_{l_z}$ at different values of $t_{tun}$ and $N_s$. Each plot shows the variation in $\Delta_{l_z}$ with the magnetic field angle $\theta$ and the layer index $l_z$. These results support the main finding \textit{i.e.} there are two distinct ranges of $\theta$ \textit{viz.}  the suppressed (OFF) and the enhanced (ON) superconducting states. The profile of $\Delta_{l_z}$ remains nearly unchanged within the angle range $100^{\circ} \leq \theta \leq 260^{\circ}$ (at $N_s$=$15$ and $t_{tun}$=$t$). Though there is a profile of the superconducting gap inside the SC slab, the entire SC slab is superconducting, implying a zero-resistance state inside it. The switching effect is, therefore, achievable at different $t_{tun}$, the smallest being $t_{tun}$=$0.5t$ at which the switching in the superconducting state appears when $N_s$ is equal to or below $N_s$=$9$. We further assert that the switching effect at even smaller $t_{tun}$ than $t_{tun}$=$0.5t$ can be achieved by concomitantly reducing the SC slab thickness $N_s$ because $t_{tun}$ relates with the length scale of the proximity effect.

The coexistence of the spin polarization and the superconductivity is observed at metallic point contacts in TaAs~\cite{Aggarwal_NComm2017}. Our analysis on the lattice matching at different WSM/SC interfaces and experimental findings of a strong interface coupling suggest that the switching effect can be tested at Nb/TaP, Nb/NbP, In/NbP and In/TaP interfaces~\cite{SM,Grabecki_PRB2020,Bachmanne1602983}. The strength of the external magnetic field is smaller than the critical field of the superconductor ($\approx$$0.4$ T for Nb~\cite{Butler_PRL1980}). The switching effect can be further tuned by the chemical potential in the SC slab~\cite{SM}.

The potential advantages of the proposed WSM-based switching mechanism over the existing mechanisms are the following. In the existing FM-based switching devices, the presence of N\'eel domains leads to the coexistence of both standard ($R_p$>$R_{ap}$) and inverse ($R_p$<$R_{ap}$) switching effects~\cite{Zhu_PRL2009}. In the proposed WSM-based switching mechanism, the unanticipated coexistence of the standard and the inverse switching effects is not possible because the spin textures on the Fermi arcs, which are fixed in a given WSM, are robust due to strong spin-momentum locking and, therefore, they are expected to be free from any domain structure. Also, the strong spin-momentum locking and the weak Coulomb interaction in WSMs such as TaAs or NbAs ensure a sharp switching, free from the Coulomb-drag or magneto-Coulomb effect that causes trouble in existing FM-based devices~\cite{Vignale_PRB2005,PhysRevB.73.220406,Hiep_JAP2017}.

To conclude, we predict a new magnetic-switching mechanism in WSM/SC/WSM heterostructures, which employs the strong spin-momentum locking and the surface spin polarization of the WSMs. We showed that a sharp switching phenomenon arises due to the interplay between the WSM surface magnetization, superconductivity, and the external magnetic field. The switching effect is testable using a WSM with either broken time-reversal symmetry or broken both inversion and time-reversal symmetries and has the potential for novel spin-valve applications.

\section*{Acknowledgements} 
\vspace{-4mm}
N.M. acknowledges discussions with A. Matos-Abiague. This work was supported by the U.S. Department of Energy (DOE), Office of Science, Basic Energy Sciences (BES), Materials Sciences and Engineering Division.


%

\newpage
\onecolumngrid 
\vspace{6mm}
\begin{center}
\textbf{\large Supplemental Material on\\ 
Magnetic Switching in Weyl Semimetal-Superconductor Heterostructures}\\
\end{center}
\vspace{4mm}
\setcounter{equation}{0}
\setcounter{figure}{0}
\setcounter{table}{0}
\setcounter{page}{1}
\makeatletter
\renewcommand{\theequation}{E\arabic{equation}}
\renewcommand{\thefigure}{S\arabic{figure}}
\renewcommand{\bibnumfmt}[1]{[R#1]}
\renewcommand{\citenumfont}[1]{R#1}


\noindent {\bf Contents:}
\begin{enumerate}[itemsep=0.5mm]
\item Hamiltonian for the middle superconductor
\item Hamiltonian for the trilayer heterostructure
\item Calculation of the order parameters
\item External control on the switching effect
\item Absence of the switching effect at zero WSM/SC coupling strength
\item Proximity effect in a WSM/SC bilayer
\item Lattice matching at WSM/SC interfaces\\
\end{enumerate}

\vspace{1em}

\noindent {\bf \\1. Hamiltonian for the middle superconductor:\\}
\indent In Eq.~(2) of the main text, we used a mean-field Hamiltonian for the middle SC layer. Below we describe the standard mean-field decomposition of an interacting Hamiltonian to obtain a non-interacting one which is quadratic in fermionic operators and used in Eq.~(3). The full Hamiltonian describing the attractive pairing interaction inside the SC slab is written as
\begin{align}
\mathcal{H}_{s}=& \sum_{\mathbf{k},\sigma} \xi_k d_{\mathbf{k},\sigma}^{\dagger} d_{\mathbf{k},\sigma} 
+ \frac{1}{N}\sum_{\mathbf{k,k'}}U_{{kk'}} d_{\mathbf{k},\uparrow}^{\dagger} d_{-\mathbf{k},\downarrow}^{\dagger}d_{-\mathbf{k'},\downarrow}d_{\mathbf{k'},\uparrow}
\end{align}
where the first term represents the kinetic energy of electrons, $\xi_k$=$-2t_s(\cos{k_x}\!+\!\cos{k_y}\!+\!\cos{k_z})\!-\!\mu_s$, $N$ is the total number of momenta in the Brillouin zone and $U_{{kk'}}$ is the strength of the boson-induced effective interaction between electrons at momenta $\mathbf{k}$ and $\mathbf{k'}$. The quartic term in the above Hamiltonian is decoupled into quadratic terms using the usual mean field theory:
$
\langle d_{\mathbf{k},\uparrow}^{\dagger} d_{-\mathbf{k},\downarrow}^{\dagger}d_{-\mathbf{k'},\downarrow}d_{\mathbf{k'},\uparrow} \rangle
\approx \langle d_{\mathbf{k},\uparrow}^{\dagger} d_{-\mathbf{k},\downarrow}^{\dagger} \rangle  d_{-\mathbf{k'},\downarrow}d_{\mathbf{k'},\uparrow} +
 d_{\mathbf{k},\uparrow}^{\dagger} d_{-\mathbf{k},\downarrow}^{\dagger} \langle  d_{-\mathbf{k'},\downarrow}d_{\mathbf{k'},\uparrow} \rangle -
 \langle d_{\mathbf{k},\uparrow}^{\dagger} d_{-\mathbf{k},\downarrow}^{\dagger} \rangle \langle  d_{-\mathbf{k'},\downarrow}d_{\mathbf{k'},\uparrow} \rangle.
$\\
With the mean-field order parameter $\Delta_{k}=(1/N)\sum_{k'} \langle d_{\mathbf{k'},\uparrow} d_{-\mathbf{k'},\downarrow} \rangle$, the Hamiltonian $\mathcal{H}_{s}$ becomes
\begin{align}
\mathcal{H}_{s}= \sum_{\mathbf{k},\sigma} \xi_k d_{\mathbf{k},\sigma}^{\dagger} d_{\mathbf{k},\sigma} 
+  \sum_{\mathbf{k,k'}} (U_{{kk'}}\Delta_k d_{\mathbf{k},\uparrow}^{\dagger}d_{\mathbf{-k},\downarrow}^{\dagger}+H.c.) 
- \sum_{\mathbf{k,k'}}  U_{{kk'}}\Delta_{k} \langle d_{\mathbf{k},\uparrow}^{\dagger} d_{-\mathbf{k},\downarrow}^{\dagger} \rangle.
\end{align}
For pairing at zero center-of-mass momentum ($|\mathbf{k}-\mathbf{k'}|$=$0$) and isotropic $s$-wave superconductivity ($\Delta_{k}$=$\Delta_{s}$ and $U_{{kk'}}$=$U_0$), the above equation becomes the following
\begin{align}
\mathcal{H}_{s}&= \sum_{\mathbf{k},\sigma} \xi_k d_{\mathbf{k},\sigma}^{\dagger} d_{\mathbf{k},\sigma} 
+  \sum_{\mathbf{k}} (U_0\Delta_s d_{\mathbf{k},\uparrow}^{\dagger}d_{\mathbf{-k},\downarrow}^{\dagger}+H.c.) +  NU_0\Delta_{s}^{2}
\end{align}
which is the same Hamiltonian expressed in Eq.~(3) of the main text. The above Hamiltonian is converted into a slab Hamiltonian as described below.\\

\noindent {\bf 2. Hamiltonian for the trilayer heterostructure:\\}
\indent To derive the Hamiltonian of the WSM/SC/WSM heterostructure, we first obtain the Hamiltonian for the WSM and the SC slabs separately and then couple the three Hamiltonian matrices via a tunneling Hamiltonian matrix. In the slab geometry, we consider open boundary conditions along the $z$ direction and periodic boundary conditions along the $x$ and $y$ directions~\cite{Mohanta_SciRep2017}. Below, we present the derivation for the Hamiltonian of the WSM slab and similar procedure is used to obtain the Hamiltonian for the SC slab.\\
\indent The choice of the boundary conditions allows us to perform a partial Fourier transformation of the fermionic operators as $\Psi_{\mathbf{k}}$=$(1/N_{w})e^{ik_zl_z}\Psi_{k_x,k_y,l_z}$, where $N_w$ is the total number of layers inside the WSM slab along the $z$ direction and $l_z$ is the real-space layer index. 
To describe the proximity-induced superconductivity in the WSM, we first express $\mathcal{H}_{w}$ in the Nambu-spinor basis $\Psi_{\mathbf{k}}$=$\big( c_{\mathbf{k},\uparrow}, c_{\mathbf{k},\downarrow}, c_{\mathbf{-k},\uparrow}^{\dagger}, c_{\mathbf{-k},\downarrow}^{\dagger} \big)^{T}$ as
\begin{align}
\mathcal{H}_{w}^{sc}\!=\!\sum_{\mathbf{k}}
\begin{pmatrix}
\mathcal{H}_{w} & 0\\
0  &  -\mathcal{H}^{*} _{w}
\end{pmatrix},
\end{align}
which is then converted to a slab Hamiltonian by writing it as
\begin{align}
\mathcal{H}_{w, slab}^{sc}&=\sum_{\mathbf{k}} \Psi_{\mathbf{k}}^{\dagger} \mathcal{H}_{w}^{sc} \Psi_{\mathbf{k}}  \nonumber \\
&=\frac{1}{N_{w}^{2}} \sum_{\mathbf{k}} \sum_{l_z,l_z^{\prime}} e^{-i k_z l_z} \Psi_{k_x,k_y,l_z}^{\dagger} \mathcal{H}_{w}^{sc} e^{i k_z l_z^{\prime}} \Psi_{k_x,k_y,l_z^{\prime}}.
\label{H_w_sc_slab}
\end{align}
With the exponential forms of the sine and cosine functions and the identities $\sum_{k_z}e^{ik_z(l_z-l_z^{\prime}-1)}$=$N_w\delta_{z,l_z^{\prime}+1}$, $\sum_{k_z}e^{ik_z(l_z-l_z^{\prime}+1)}$=$N_w\delta_{z,l_z^{\prime}-1}$, Eq.~(\ref{H_w_sc_slab}) becomes 
\begin{align}
\mathcal{H}_{w, slab}^{sc}=\frac{1}{N_w} \sum_{\mathbf{k}^{\parallel}} \Big( \sum_{l_z=1}^{N_w} \Psi_{\mathbf{k}^{\parallel},l_z}^{\dagger} \mathcal{H}_0(\mathbf{k}^{\parallel}) \Psi_{\mathbf{k}^{\parallel},l_z} 
+ \sum_{l_z=1}^{N_w-1} \big[ \Psi_{\mathbf{k}^{\parallel},l_z}^{\dagger} \mathcal{H}_1 \Psi_{\mathbf{k}^{\parallel},l_z+1} +H.c.\big] \Big),
\end{align}
where $\mathbf{k}^{\parallel}$$\equiv$$(k_x,k_y)$, $\mathcal{H}_{0}(\mathbf{k}^{\parallel})$ is the Hamiltonian for a single layer in the $k_x$-$k_y$ plane within the WSM slab and is given by
\begin{align}
\mathcal{H}_{0}=
\begin{pmatrix}
\mathcal{H}_{w}^{\parallel} & 0  \\
0  & -\mathcal{H}_{w}^{\parallel,*}
\end{pmatrix},
\end{align}
with $\mathcal{H}_{w}^{\parallel}=-[ m(2-\cos{k_y})+2t_x(\cos{k_x} -\cos{k_0}) ] \sigma_{x} -(2t \sin{k_y}) \sigma_{y}$. The interlayer coupling, described by Hamiltonian $\mathcal{H}_{1}$, is given by
\begin{align}
\mathcal{H}_{1}=
\begin{pmatrix}
\mathcal{H}_{w}^{\perp} & 0 \\
0 & - \mathcal{H}_{w}^{\perp,*} \end{pmatrix},
\end{align}
where $\mathcal{H}_{w}^{\perp}=m/2 \sigma_x +it \sigma_z$.
Following the same prescription as above, the Hamiltonian $\mathcal{H}_{s}(\mathbf{k})$ (Eq.~(2) in main text) is transformed into a slab Hamiltonian.\\
\indent In the main text, we consider a geometry in which the top and the bottom WSMs have opposite alignments of the surface Fermi arcs. This is achieved by reversing the order of the layers along the $z$ direction in the slab Hamiltonian matrices for the top and the bottom WSM slabs. This type of alignment can also be realized in the experiment by performing a $180^{\circ}$ rotation of the top WSM slab sample in the $x$-$y$ plane after locating the direction of the surface spin polarization.\\

\noindent {\bf 3. Calculation of the order parameters:\\}
\indent The superconducting order parameters, the pairing amplitudes at each layer $l_z$ of the heterostructure, is obtained by the self-consistent calculation, described below. The pairing amplitude $\Delta_{s,l_z}$=$\langle d_{\mathbf{k^{\parallel}},l_z,\uparrow} d_{\mathbf{-k^{\parallel}},l_z,\downarrow} \rangle$, at layer $l_z$ inside the SC, is obtained by minimizing the free energy of the heterostructure Hamiltonian. To obtain the free energy, we first write down the partition function as below
\begin{align}
\mathcal{Z}=Tr[e^{-\beta \mathcal{H}_{het}}],
\end{align}
where $\mathcal{H}_{het}$ is the full Hamiltonian of the WSM/SC/WSM heterostructure, $\beta$=$1/k_{B}T$, $k_B$ is the Boltzmann constant and $T$ is the temperature. The grand canonical free energy  $\mathcal{F} =-1/\beta {\rm ln} \mathcal{Z}$ is given by 
\begin{align}
\mathcal{F} = -\frac{1}{N_k\beta} \sum_{\mathbf{k^{\parallel}},\lambda, l_z} {\rm ln}(1+e^{-\beta E_{\mathbf{k^{\parallel}} \lambda}}) -   \sum_{l_z} U_0N_k \Delta_{s,l_z}^{2} ,
\end{align}
where $N_k$ is the total number of momenta in the two-dimensional Brillouin zone in each layer and $E_{\mathbf{k^{\parallel}} \lambda}$ is the $ \lambda^{th}$ eigen value of the heterostructure at momentum $\mathbf{k^{\parallel}}$.
The set of self-consistency equations for $\Delta_{s,l_z}$ is given by setting $\partial \mathcal{F}/\partial \Delta_{s,l_z}$=$0$
\begin{align}
\Delta_{s,l_z}=\frac{1}{2U_0N_k} \sum_{\mathbf{k^{\parallel}},\lambda} \frac{\partial E_{\mathbf{k^{\parallel}} \lambda}}{\partial \Delta_{s,l_z}} \tanh\Big( \frac{\beta E_{\mathbf{k{\parallel}} \lambda} }{2} \Big).
\label{gap_eq}
\end{align}
We start with a finite initial value of $\Delta_{s,l_z}$ and update them iteratively until self-consistency is achieved at each layer $l_z$. The pairing amplitude $\Delta_{w,l_z}$=$\langle c_{\mathbf{k^{\parallel}},l_z,\uparrow} c_{\mathbf{-k^{\parallel}},l_z,\downarrow} \rangle$ at layer $l_z$ inside the WSM is calculated using the four-component spinor wave functions, extracted from the converged wave function of $\mathcal{H}_{het}$, and using the usual Bogoliubov-Valatin transformation $c_{\mathbf{k^{\parallel}},l_z,\sigma}$=$\sum_{\lambda}u_{\mathbf{k^{\parallel}},l_z,\sigma}^{\lambda}\gamma_{_{\lambda}}$+$v_{\mathbf{k^{\parallel}},l_z,\sigma}^{\lambda *}\gamma_{_{\lambda}}^{\dagger}$, where $u_{\mathbf{k^{\parallel}},l_z,\sigma}^{\lambda}\gamma_{_{\lambda}}$ ($v_{\mathbf{k^{\parallel}},l_z,\sigma}^{\lambda}\gamma_{_{\lambda}}$) is the quasiparticle (quasihole) amplitude and $\gamma_{_{\lambda}}$ ($\gamma_{_{\lambda}}^{\dagger}$) is the fermionic annihilation (creation) operator. The pairing amplitude $\Delta_{w,l_z}$ is then obtained using
\begin{align}
\Delta_{w,l_z}=\sum_{\mathbf{k^{\parallel}},\lambda} \Big[ u_{\mathbf{k^{\parallel}},l_z,\uparrow}^{\lambda} v_{\mathbf{k^{\parallel}},l_z,\downarrow}^{\lambda *} (1-f(E_{\mathbf{k^{\parallel}} \lambda})) 
+u_{\mathbf{k^{\parallel}},l_z,\downarrow}^{\lambda} v_{\mathbf{k^{\parallel}},l_z,\uparrow}^{\lambda *} f(E_{\mathbf{k^{\parallel}} \lambda}) \Big],
\end{align}
where $f(E_{\mathbf{k^{\parallel}} \lambda})$ is the Fermi function at temperature $T$ and energy $E_{\mathbf{k^{\parallel}} \lambda}$. \\
\indent The magnetization components at each layer $l_z$ and momentum $\mathbf{k}^{\parallel}$ is obtained by taking the expectation values of the Pauli matrices in the particle-hole space, $\tau_{\alpha}$ (where $\alpha$=$x,y,z$), on the calculated wave function $\Psi_{\mathbf{k}^{\parallel},l_z}$ as follows
\begin{align}
m_{\mathbf{k}^{\parallel},l_z,\alpha} = \frac{\hbar}{2} \Big \langle \Psi_{\mathbf{k}^{\parallel},l_z} \Big |  \tau_{\alpha} \Big |  \Psi_{\mathbf{k}^{\parallel},l_z} \Big \rangle,
\end{align}
where  $\Psi_{\mathbf{k}^{\parallel},l_z}$=$\big( c_{\mathbf{k}^{\parallel},l_z,\uparrow}, c_{\mathbf{k}^{\parallel},l_z,\downarrow} c_{-\mathbf{k}^{\parallel},l_z,\downarrow}^{\dagger}, c_{-\mathbf{k}^{\parallel},l_z,\uparrow}^{\dagger} \big)$ for the WSM slab, and  similar for the SC slab with $c_{\mathbf{k}^{\parallel},l_z,\sigma}$ (here $\sigma$=$\uparrow,\downarrow$) operators replaced by $d_{\mathbf{k}^{\parallel},l_z,\sigma}$ operators.\\

\begin{figure}[h]
\begin{center}
\vspace{5mm}
\epsfig{file=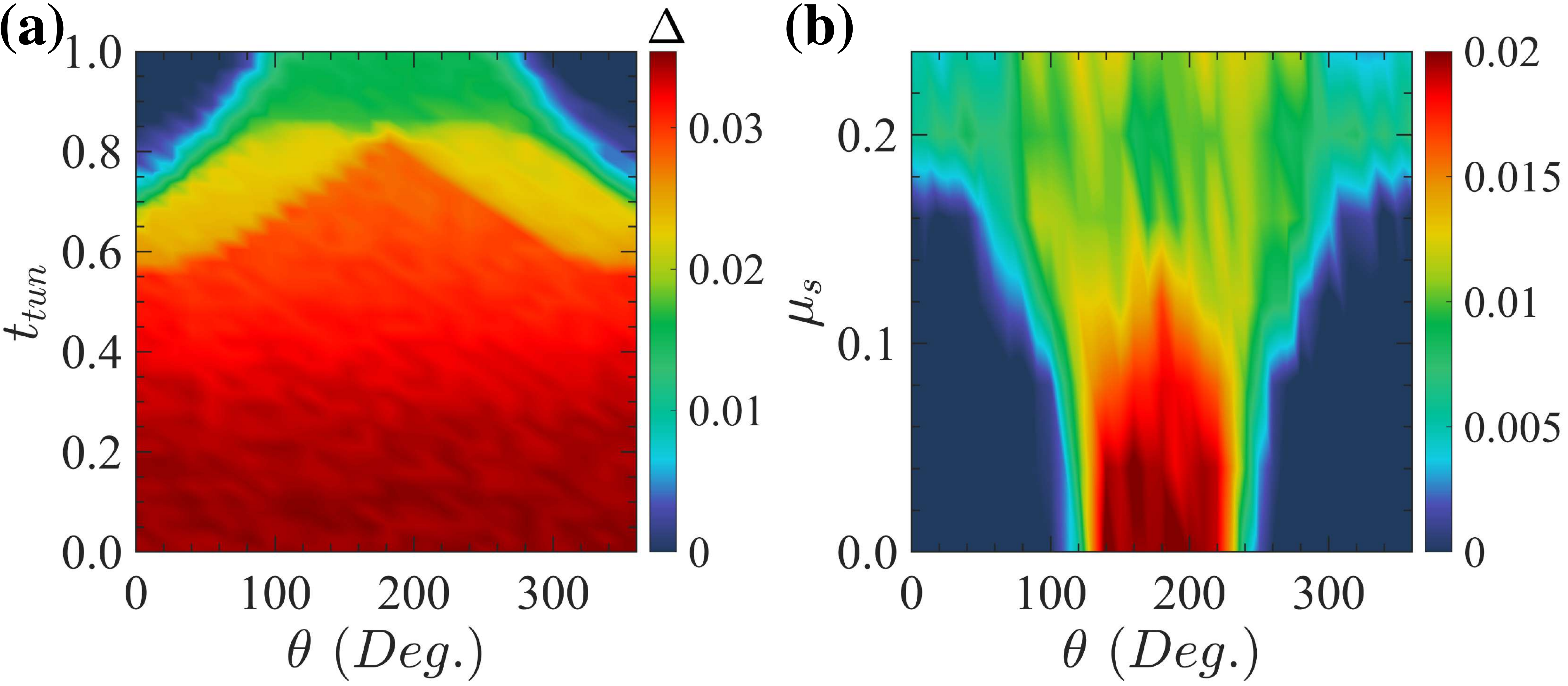,trim=0.0in 0.2in 0.0in 0.1in,clip=false, width=85mm}
\caption{Variation of the mean superconducting pairing amplitude $\Delta$ of the trilayer heterostructure with magnetic-field angle $\theta$ and (a) the WSM-SC tunneling amplitude $t_{tun}$ and (b) the chemical potential $\mu_s$ in the SC slab. Other parameters, unless otherwise stated, are the same as in Fig.~(2) of the main text.}
\label{control}
\vspace{-4mm}
\end{center}
\end{figure}

\noindent {\bf 4. External control on the switching effect:\\}
\indent The magnetic-field induced switching effect, which is discussed in the main text, can be controlled by a number of ways. In Fig.~\ref{control}, we show the modulation of the mean superconducting pairing amplitude by the WSM-SC tunneling amplitude $t_{tun}$ and by the chemical potential $\mu_s$ in the SC. Fig.~\ref{control}(b) indicates that the switching effect is tunable by $\mu_s$ which can be modulated externally by a gate voltage.\\

\begin{figure}[h]
\begin{center}
\epsfig{file=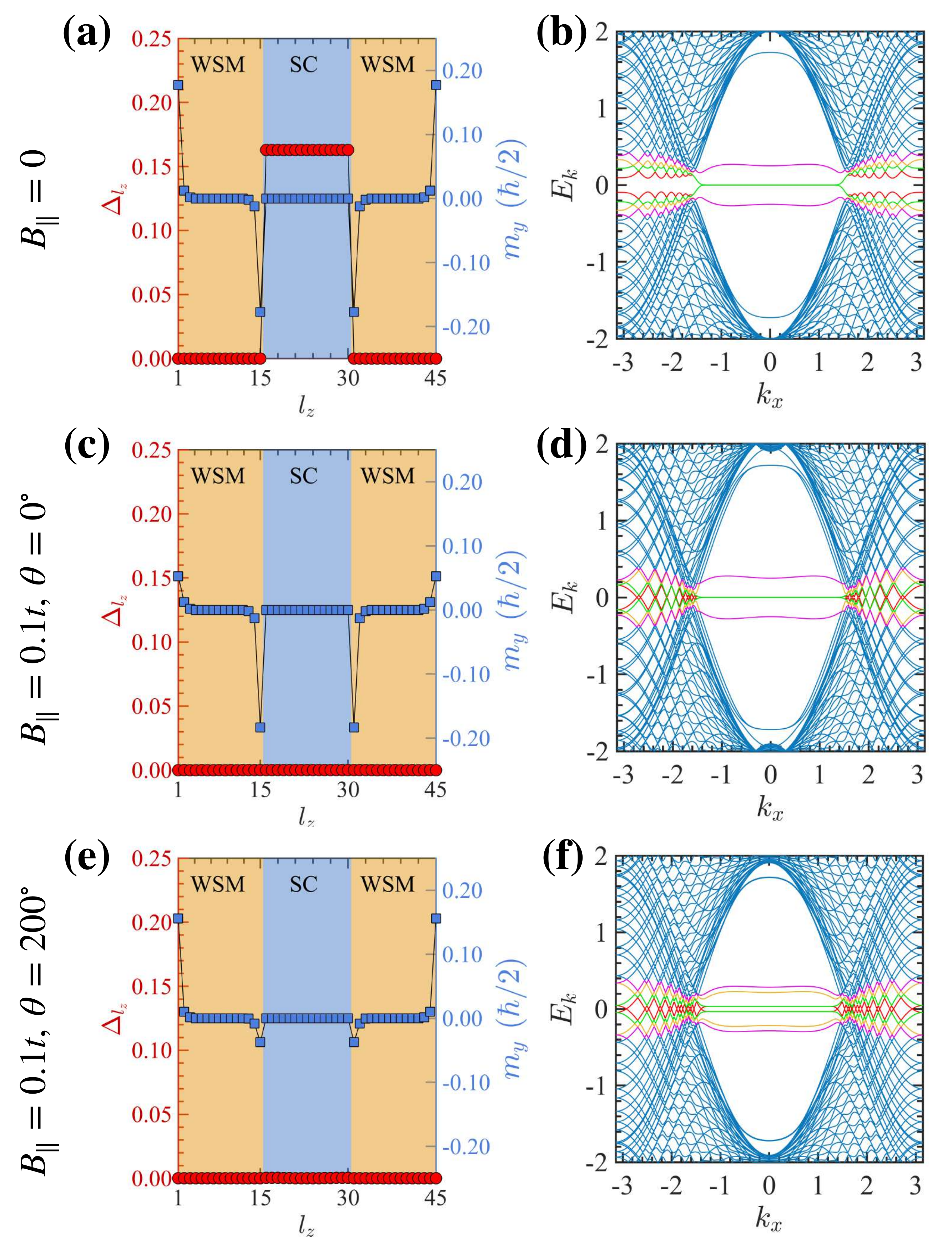,trim=0.0in 0.0in 0.0in 0.0in,clip=false, width=85mm}
\caption{(a),(c),(e) Variation of the superconducting pairing amplitude $\Delta_{l_z}$ (left vertical axis) and  $y$ component of magnetization $m_y$ (right vertical axis) with the layer index $l_z$ at (a) magnetic field $B_{\parallel}$=$0$, (c) $B_{\parallel}$=$0.1t$, $\theta$=$0^{\circ}$ and (e) $B_{\parallel}$=$0.1t$, $\theta$=$200^{\circ}$. (b), (d) and (f) show the corresponding energy spectrum of the WSM/SC/WSM heterostructure at $k_y$=$0$. The energy levels in red, green, yellow and magenta originate primarily from the surfaces of the WSMs. Parameters are the same as in Fig.~2 of the main text, except $t_{tun}$=$0$.}
\label{t_tun0}
\vspace{-8mm}
\end{center}
\end{figure}

\noindent {\bf 5. Absence of the switching effect at zero WSM/SC coupling:\\}
\indent To understand the role of the WSM/SC interface on the observed switching effect, here we consider the case in which the WSM and the SC slabs are not connected \textit{i.e.} $t_{tun}$=$0$.  The layer-resolved pairing amplitude $\Delta_{lz}$, and the dominant magnetization component $m_y$ are shown in Fig.~\ref{t_tun0}(a), (c) and (e), at different field scenarios and the corresponding spectrum in Fig.~\ref{t_tun0}(b), (d) and (f).  The magnetization inside the WSM behaves in a similar way as in the case $t_{tun}$$\neq$$0$, described in the main text \textit{i.e.} the surface magnetization decreases when the magnetic field is applied opposite to the spin polarization direction. With $t_{tun}$$\neq$$0$, both the WSM surface magnetization and the magnetic field jointly suppress the superconductivity. In the case $t_{tun}$=$0$, a higher magnetic field is required to fully suppress the superconductivity. However, this field is equal to or larger than the critical field $B_{\parallel}^c$$\approx$$0.3t$  for superconductivity in the SC slab and, therefore, a reappearance of superconductivity is not possible. Furthermore, the spectra in Fig.~\ref{t_tun0}(b), (d) and (f) do not reveal the reappearance of the spectral gap outside the flat-band region (\textit{i.e.} $|k_x|$$>$$k_0$), as found in the case with $t_{tun}$$\neq$$0$. The results highlight the importance of a coupled WSM/SC interface behind the discussed switching effect.\\

\begin{figure}[h]
\begin{center}
\epsfig{file=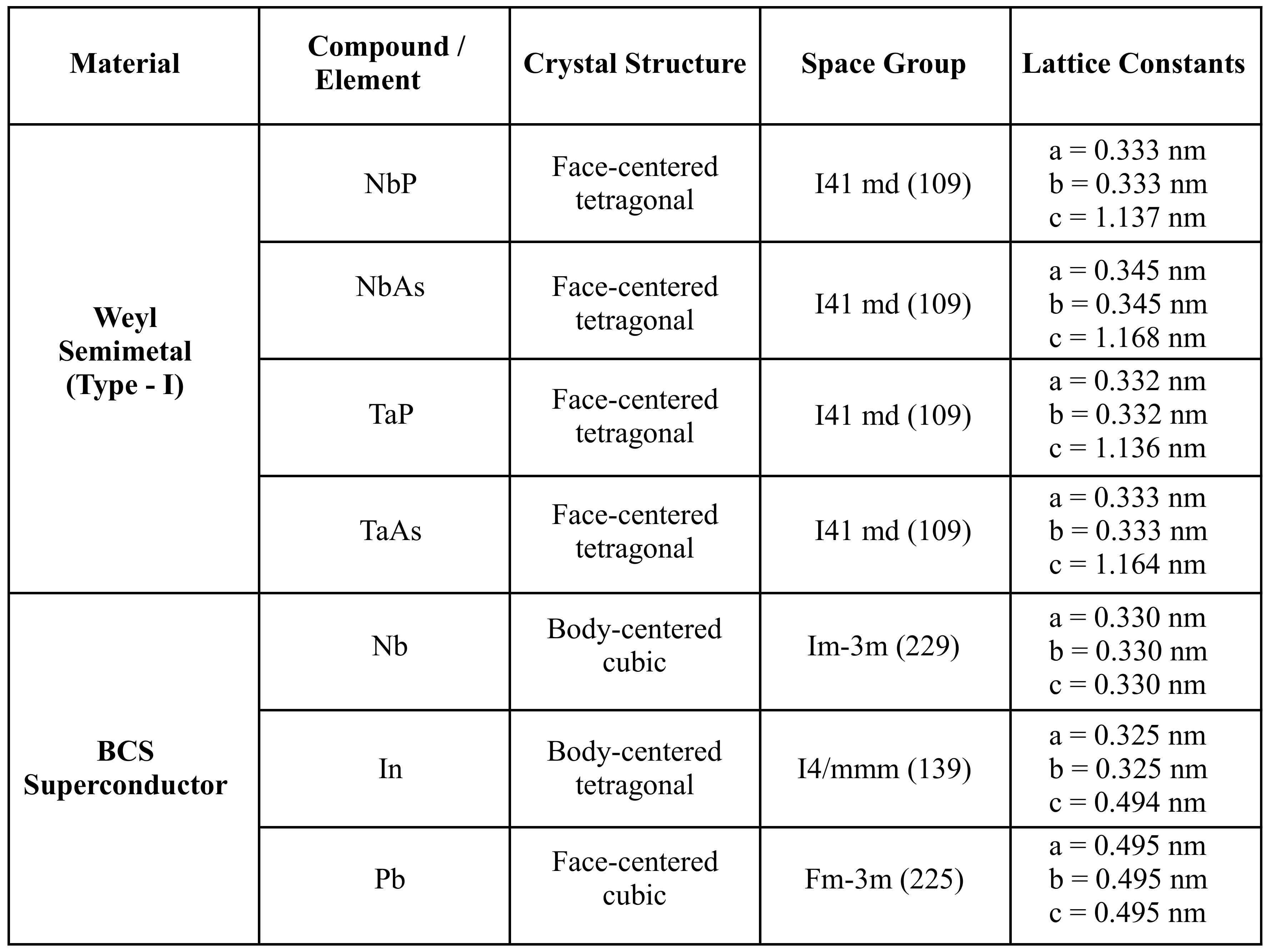,trim=0.0in 0.0in 0.0in 0.0in,clip=false, width=85mm}
\caption{Structural information of the WSMs and the SCs which have been demonstrated in Ref.~\onlinecite{Grabecki_PRB2020} to form a robust WSM/SC interface, leading to a strong proximity effect. The data presented in above table were taken from references~\cite{Sun_PRB2015,Villars2016:sm_isp_sd_1215974,Villars2016:sm_isp_sd_0451186,Villars2016:sm_isp_sd_1003520,Villars2016:sm_isp_sd_0529060,Villars2016:sm_isp_sd_0260306,Villars2016:sm_isp_sd_0452697}.}
\label{table1}
\vspace{0mm}
\end{center}
\end{figure}

\begin{figure}[h]
\begin{center}
\epsfig{file=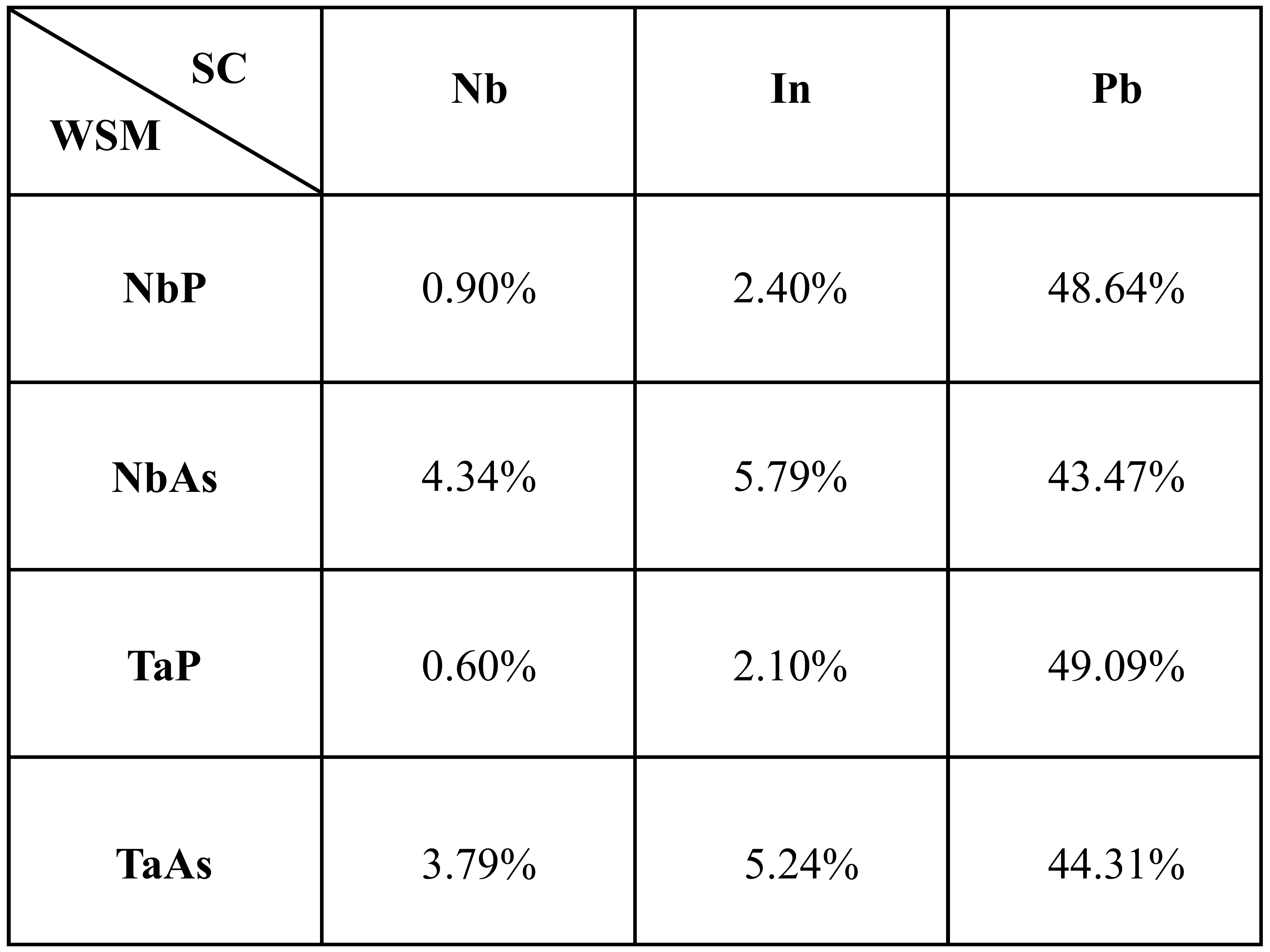,trim=0.0in 0.0in 0.0in 0.0in,clip=false, width=65mm}
\caption{Percentage lattice mismatch at a WSM/SC interface, obtained via  $(a_{_{WSM}}-a_{_{SC}})/a_{_{WSM}}\times100\%$, where $a_{_{WSM}}$ and  $a_{_{SC}}$ are the lattice constants of the WSM and the SC, respectively.}
\label{table2}
\vspace{0mm}
\end{center}
\end{figure}

\noindent {\bf 6. Lattice matching at WSM/SC interfaces:\\}
\indent Here, we discuss the possible WSM/SC interfaces (see Fig.~\ref{table1} and Fig.~\ref{table2}) that could lead to a strong proximity coupling between the two materials. Recent experiments suggest that the four type-I WSM compounds NbP, NbAs, TaP and TaAs form good interfaces with superconducting elements Nb, In, and Pb~\cite{Grabecki_PRB2020, Bachmanne1602983}. From the analysis below of lattices that may match from the crystallographic perspective, we find that Nb/TaP, Nb/NbP, In/NbP and In/TaP are good candidates where a strong proximity coupling can be expected. In fact, Grabechi et al. already reported the formation of a very strong SC-WSM contact in an In-NbP junction by looking at the superconducting proximity effect~\cite{Grabecki_PRB2020}. These WSM/SC interfaces are, therefore, promising platforms to test the proposed switching effect and search for unconventional phenomena which can appear because of the magnetic proximity effect between WSMs and SCs.\\
\vspace{5mm}

\begin{figure}[h]
\begin{center}
\epsfig{file=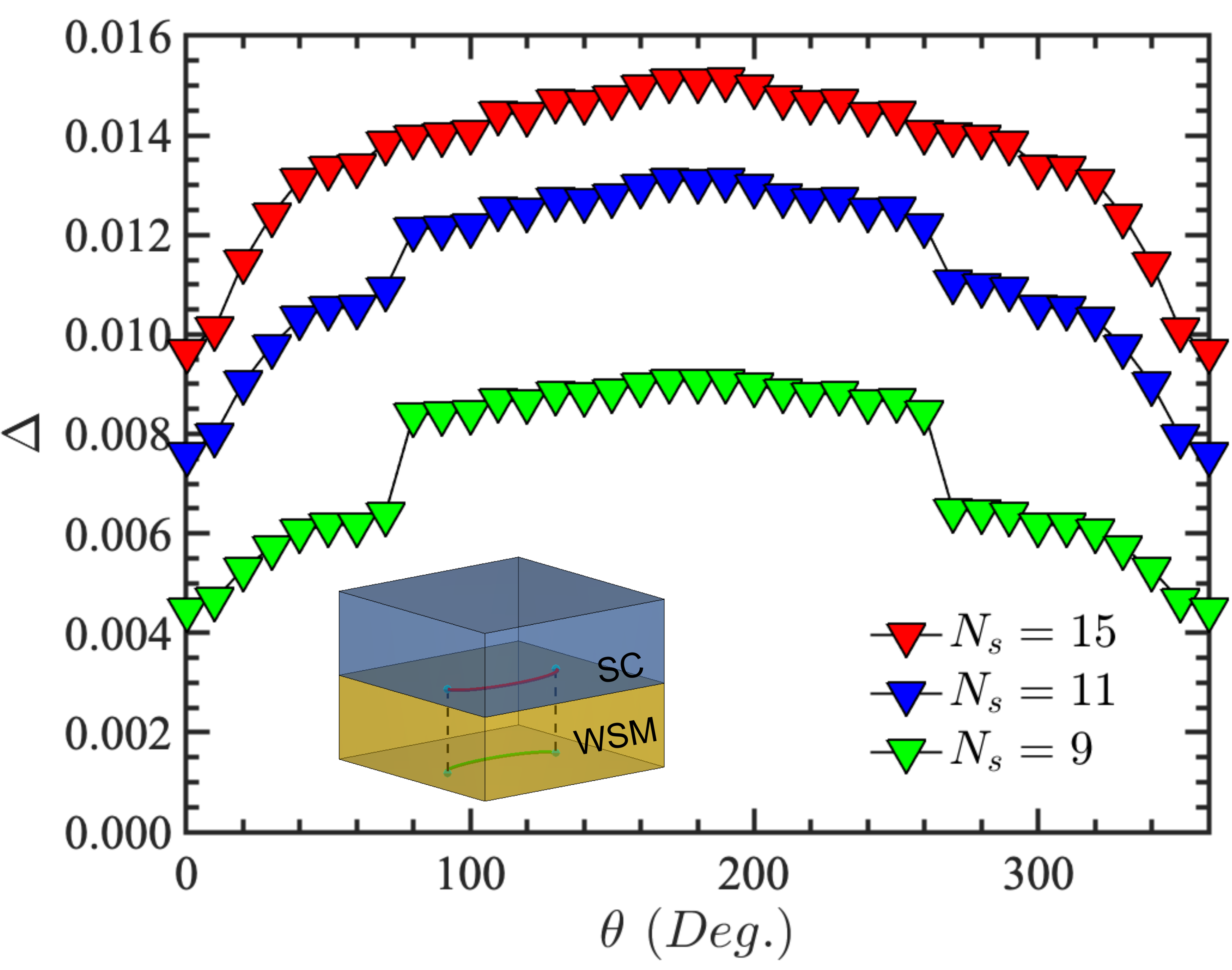,trim=0.0in 0.0in 0.0in 0.0in,clip=false, width=80mm}
\caption{Variation of the average superconducting pairing amplitude $\Delta$ with the magnetic-field angle $\theta$ in a WSM/SC bilayer at different thicknesses of the SC layer. Parameters are the same as in Fig.~2 of the main text.}
\label{bilayer}
\vspace{-8mm}
\end{center}
\end{figure}

\noindent {\bf 7. Proximity effect in a WSM/SC bilayer:\\}
\indent In a WSM/SC bilayer, the superconducting pairing amplitude is reduced when the magnetic field is applied along a particular direction, as shown in Fig.~\ref{bilayer}. This indicates that the magnetic proximity effect on superconductivity is present in the bilayer also. However, the proximity effect is much stronger in the trilayer case leading to the complete switching of the superconducting pairing amplitude. 


%

\end{document}